\documentclass{vgtc}                          % final (conference style)
\graphicspath{{figures/}{pictures/}{images/}{./}} % where to search for the images

\usepackage{times}                     % we use Times as the main font
\usepackage[T1]{fontenc} 
\usepackage{ragged2e}
\usepackage{tabu}                      % only used for the table example
\usepackage{booktabs}                  % only used for the table example
\usepackage{lipsum}                    % used to generate placeholder text
\usepackage{mwe}                       % used to generate placeholder figures

\usepackage{mathptmx}                  % use matching math font

\usepackage{amsmath}
\onlineid{0}

\vgtccategory{Research}

\vgtcinsertpkg

\title{Shape Shifters: Does Body Shape Change the Perception of Small-Scale Crowd Motions?}

\author{Bharat Vyas\thanks{e-mail: vyasb@tcd.ie}\\ %
        \scriptsize Trinity College Dublin %
\and Carol O'Sullivan\thanks{e-mail: carol.osullivan@tcd.ie}\\ %
     \scriptsize Trinity College Dublin }

\teaser{
 \includegraphics[width=0.9\linewidth]{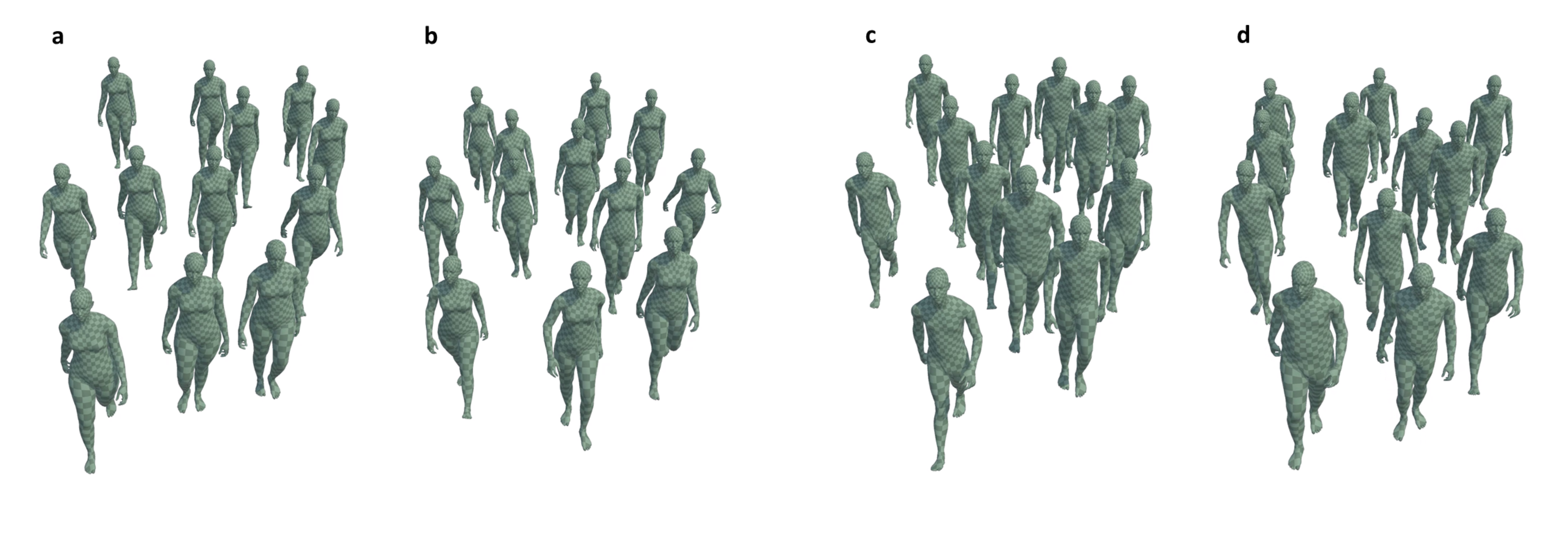}
 \centering
  \caption{Example small-scale crowds of virtual avatars consisting of:  
\textbf{a)} Female avatars with cloned motions (out-of-step) and identical body shapes, 
\textbf{b)} Female avatars with unique motions and different body shapes, 
\textbf{c)} Male avatars with cloned motions (out-of-step) and different body shapes,  
\textbf{d)} Male avatars with unique motions and different body shapes.  }
\label{fig:teaser}
\vspace{-0.1cm}
}

\abstract{
    The animation of realistic virtual avatars in crowd scenarios is an important element of immersive virtual environments. However, achieving this realism requires attention to multiple factors, such as their visual appearance and motion cues. We investigated how body shape diversity influences the perception of motion clones in virtual crowds. A physics-based model was used to simulate virtual avatars in a small-scale crowd of size twelve. Participants viewed side-by-side video clips of these virtual crowds: one featuring all unique motions (Baseline) and the other containing motion clones (i.e., the same motion used to animate two or more avatars in the crowd). We also varied the levels of body shape and motion diversity. Our findings revealed that body shape diversity did not influence participants' ratings of motion clone detection, and motion variety had a greater impact on their perception of the crowd. Further research is needed to investigate how other visual factors interact with motion in order to enhance the perception of virtual crowd realism.
} % end of abstract


\begin{document}


\firstsection{Introduction}

\maketitle

The presence of realistic virtual avatars is a common feature of virtual environments, such as virtual reality (VR) games and simulations \cite{magnenat2005handbook}. However, achieving realism becomes more challenging when multiple avatars are present in group or crowd settings. These scenarios are frequently encountered in everyday life, where small and large groups of people stroll in parks or city streets. With advances in computer graphics and visual effects, small- and large-scale crowds are frequently recreated in virtual environments for use in movies and interactive applications. 

However, a balance between realism and system performance is needed. 
Researchers have developed techniques to enhance virtual crowd realism while ensuring computational efficiency. These include diversifying appearance through body shape \cite{shi2017shape,silva2019crowd}, texture \cite{tecchia2002image}, and colour \cite{ulicny2004crowdbrush}, as well as improving motion realism using motion blending \cite{huang2010motion} and secondary motion \cite{hoyet2016perceptual}.
% Researchers have explored various techniques to enhance the realism of virtual crowds while maintaining computational efficiency. This involves creating diversity in appearance such as body shape \cite{shi2017shape,silva2019crowd}, texture \cite{tecchia2002image}, colour \cite{ulicny2004crowdbrush}, and accessories \cite{maim2009yaq}. In addition, motion realism has been enhanced through methods like motion blending \cite{huang2010motion} and secondary motion \cite{hoyet2016perceptual}.

In real world crowd scenarios, each individual has their own unique appearance and movements. However, achieving this level of variety across all avatars in a virtual crowd is computationally expensive and inefficient. It is therefore important to determine what level of variety is needed for a virtual crowd to appear realistic and natural. 
% Extensive research has been conducted to develop techniques that introduce variety into crowd scenarios, particularly in crowd rendering \cite{beacco2016survey,dobbyn2005geopostors}. 
One key aspect of crowd realism is the variety of 3D models used to depict virtual avatars. As noted by Shi et al. \cite{shi2017shape}, shape and size are among the most significant and visually impactful features of any crowd representation. Previous research by McDonnell et al. \cite{mcdonnell2008clone} examined the identifiability of appearance clones while varying the textures and colours of clothing. In this paper, we explore the additional effect of body shape and how it interacts with motion to influence the perception of small-scale crowds.

Regarding motion variety, Pražák et al. \cite{pravzak2011perceiving} previously reported that a limited number of motions were sufficient to animate a realistic crowd. They found that a crowd consisting of multiple motion clones (in their case, three unique motions replicated evenly throughout the crowd) was perceived to be as varied as one in which all motions were unique. Building on these findings, we examine the effect of body shape variety on the perception of motion clones. 

Our small-scale crowds are created using physics-based virtual avatars, who follow kinematic motion but are actually driven by dynamic forces and torques to change position and joint angles and for simulation in general. These characters are rarely seen in crowd scenarios because of modelling complexity and computational demands. However, we argue that these characters could be a useful integration in such scenarios, especially small-scale crowds, as they can dynamically react to changes in the environment. For example, the shoulder push examined by Hoyet et al. \cite{hoyet2016perceptual} demonstrated that secondary motion in crowds significantly impacts the visual quality of crowd animations. Such secondary motions can be easily generated for these dynamic characters when they encounter collisions.

Our virtual crowds were created using the full-body walking motions captured by Hoyet et al. \cite{hoyet2013evaluating}. A basic control system was developed that physically controls the virtual avatars to  mimic these reference motions. The motions of 24 actors (12F/12M) and their body measurements were used to generate the individuals in the crowds. To ensure that the body shapes were sufficiently varied, the actors were selected from three different Body Mass Index (BMI) \cite{world1995physical} categories. Our baseline stimuli consisted of one male and one female crowd, each consisting of 12 virtual avatars with their own unique body shape and motion. Further stimuli were generated by varying both the number of motion clones and body shapes used (see Figure \ref{fig:teaser} for some examples). We hypothesized that: 

%To create our baseline stimuli, we created one male and one female crowd. In each case, four body shapes were selected from each of the three BMI groups, giving 12 virtual avatars with their own unique body shape and motion. We then generated different levels of motion clones (1, 2, 3, and 6 different motions) for all virtual avatars in the crowd. For different levels of motion clones, we also varied the number of body shapes (1, 3, 6, and 12 different body shapes) in the crowd, respectively. We hypothesized that: 
\vspace{-0.1cm}
\begin{itemize}
    \item \textbf{H1:} Motion clones will be more easily detected with increasing numbers of clones.
    \item \textbf{H2:} Motion clones will be less easily detected with increasing variety of body shapes in the crowd.
\end{itemize}
\vspace{-0.1cm}

Our results showed that increasing the number of motion clones did indeed improve detection rates, whereas no significant effect of body shape variety was found. The significant effect of motion variety shows that motion had a stronger impact on perceived realism compared to the appearance feature of body shape. However, observation of the results does suggest that body shape may still provide a clone masking effect, perhaps when combined with other varied appearance features.

\section{Background}

\textbf{Perception of Crowd Motion:}
The perception of avatar motion has been long studied since the early work of Johansson \cite{johansson1973visual}. Their experiment demonstrated that placing just a few point light sources on a moving person's body were enough to be perceived as human motion. Since then, numerous studies have explored motion perception in relation to gender (\cite{kozlowski1977recognizing, troje2002decomposing}), one's own motion \cite{beardsworth1981ability} and the motion of a friend \cite{kozlowski1977recognizing}. While these works explored the perception of individual virtual humans, there has also been a lot of research exploring the perception of motion in crowd settings.

McDonnell et al. \cite{mcdonnell2008clone} explored the effect of crowd variety in terms of both appearance and motion. They found that cloned motions were more easily detected than appearance clones, which they could effectively mask using colour and texture variations. In a follow-up study \cite{mcdonnell2009eye}, they used an eye-tracker to determine which body parts of virtual avatars receive the most attention. They reported that observers fixated for longer on the head and upper torso of appearance clones. In this paper, we also explore the effect of varying the appearance of crowd individuals (with respect to their body shape), but focus on motion clone detection.

Adili et al.\cite{adili2021perception} investigated the threshold at which motion clones become indistinguishable in large virtual crowds. They found that even low motion variety (e.g., 1 motion per 25 characters) achieves equivalent perceptual results to fully unique crowds.
% {\color{red}Adili et al.\cite{adili2021perception} investigated the effect of motion variety and found that more than two motions in the case of virtual crowds were perceptually equivalent. For stimuli, they randomised the outlook of the avatars and took the generally available designed avatars.
Prazak and O'Sullivan \cite{pravzak2011perceiving} created a framework where the motion from one actor could be used to animate multiple characters. They found that cloning three motions (desynchronised for each character) and distributing them evenly among the crowd characters was sufficient to create the illusion of a fully varied crowd.\\
% The three motions cloned were as good as using different motions for all the characters. 

{\raggedleft{\textbf{Body Shape $\&$ Motion:}}}
The study of virtual avatar body shapes has also been an active research topic for many years, including both the reconstruction of 3D avatar body shapes and the perception of appearance. McDonnell et al. \cite{mcdonnell2007virtual} investigated the factors influencing the perceived sex of virtual avatars and found that in the case of neutral walks, appearance plays a crucial role in determining the sex of an avatar. Similarly, Johnson and Tassinary \cite{johnson2007compatibility} concluded that both appearance and motion significantly affect perceived attractiveness.
% These findings highlight the importance of appearance, particularly body shape, as a critical factor in perception. 
%McDonnell et al. \cite{mcdonnell2009eye} further emphasized this by demonstrating that certain upper body parts, which are key to defining body shape, are especially influential in perception.
\vspace{0.2cm}

In terms of body modelling, researchers have developed statistical models using extensive collections of 3D body scans. One of the most successful early models was SCAPE, introduced by Anguelov et al. \cite{anguelov2005scape}, a deformable avatar body model. This was followed by the work from Loper et al. \cite{SMPL:2015}, which introduced a probabilistic body model that uses Principal Component Analysis (PCA) on body parameters derived from scanned real avatar data. 
% The SMPL model, with its body shape and pose parameters, allows for the easy generation and animation of digital avatars with diverse body shapes.

Shi et al. \cite{shi2017shape} demonstrated that a cost-effective and realistic crowd could be generated by creating body shape models based on anthropometric parameters. They observed that, while they could provide a wide set of realistic body shapes to populate a crowd, repetitions of very distinctive shapes in the crowd could be more easily noticed. Using shapes that were closer to the average reduced the noticeability of body clones, although overall crowd variety was also diminished.
% We explore a similar question of whether the choice of distinctive and varied body shapes also masks the motion clones. 
To determine distinctiveness, they ran a user study based on comparisons of static images, where participants were asked to identify which one of three bodies was the target body they had previously seen. In our study, we also use anthropometric measurements of real actors to create our 3D models. However, our stimuli are animations of small-scale crowds, and our task is to determine the ability to detect motion clones using different body shape combinations.

Russell et al. \cite{russell2023detection} explored the perception of inconsistencies between body shape and motion by mismatching them, finding that participants could detect these inconsistencies during walking motions but struggled with other types of movements. Similarly, Kenny et al. \cite{kenny2019perceptual} explored mismatches between body shape and motion in scenarios where virtual avatars interacted with objects in their environment. Their findings suggested that people could not identify inconsistencies in these complex interactions. Vyas et al. \cite{vyas2023exploring} verified these findings, demonstrating that observers rated stimuli as more consistent when the body shape of an avatar matched the motion of an actor within the same BMI group. Our work builds on these studies by examining whether these results hold true within the context of motion clones and crowd animation.\\

{\raggedleft{\textbf{Physics-based Virtual avatars:}}}
The interest in physically simulated characters has increased significantly in recent years. Their ability to interact naturally with their environment makes them particularly valuable in immersive virtual environments. The work by Peng et al. \cite{peng2018deepmimic} used reference motion files and deep reinforcement learning to train policies that allow physically simulated characters to mimic real-world movements. This approach gained significant attention and has since been applied in various contexts, including control strategies for two-player scenarios \cite{won2021control}, characters with varying body shapes \cite{won2019learning}, and scalable approaches for multiple characters \cite{won2020scalable}.

However, as discussed by Geijtenbeek et al. \cite{geijtenbeek2012interactive}, physics-based animation can be achieved through various other techniques beyond deep reinforcement learning. Traditional methods often involve control systems that govern the movement of a virtual avatar's limbs. For instance, Faloutsos et al. \cite{faloutsos2001composable} developed complex and realistic character animations using composable controllers, which synthesise basic actions such as arm reactions, balance, and recovery from perturbations. Alvarado et al. \cite{alvarado2022generating} combined kinematic constraints with lightweight physics to create interactive behaviours for virtual avatars. Levine and Popović \cite{levine2012physically} introduced a quasi-physical simulation method that applies non-physical forces to the character's root, optimising the character's equations of motion for the control system to animate.

Despite these advances, our work focuses on a basic feedback control system, as our primary interest lies in the perception of retargeted motion rather than in generating stylised or varied movements. This approach allowed us to maintain consistency in the animations, while harnessing the flexibility of physics-based simulation for future exploration.
\pagebreak

\section{Experiment}
We conducted a perception experiment to explore how body shape affects the detection of motion clones in a crowd scenario. For avatar animation, we used physics-based avatars because they adhere to the physical laws of nature, making them appear and behave more plausibly. We argue that such physics-based frameworks have the potential to be used in generating plausible and natural-looking crowd behaviour. In this study, we refrained from using a complex framework as the generated motion could affect the experiment result. Therefore, we used a basic feedback control system that perfectly mimics the reference motion and outputs a physics-based character. In this way, we could focus more on the perception of small-scale crowds and not on the complexity of the framework.

\begin{figure}[t]
  \centering
  \includegraphics[width=1.0\linewidth]{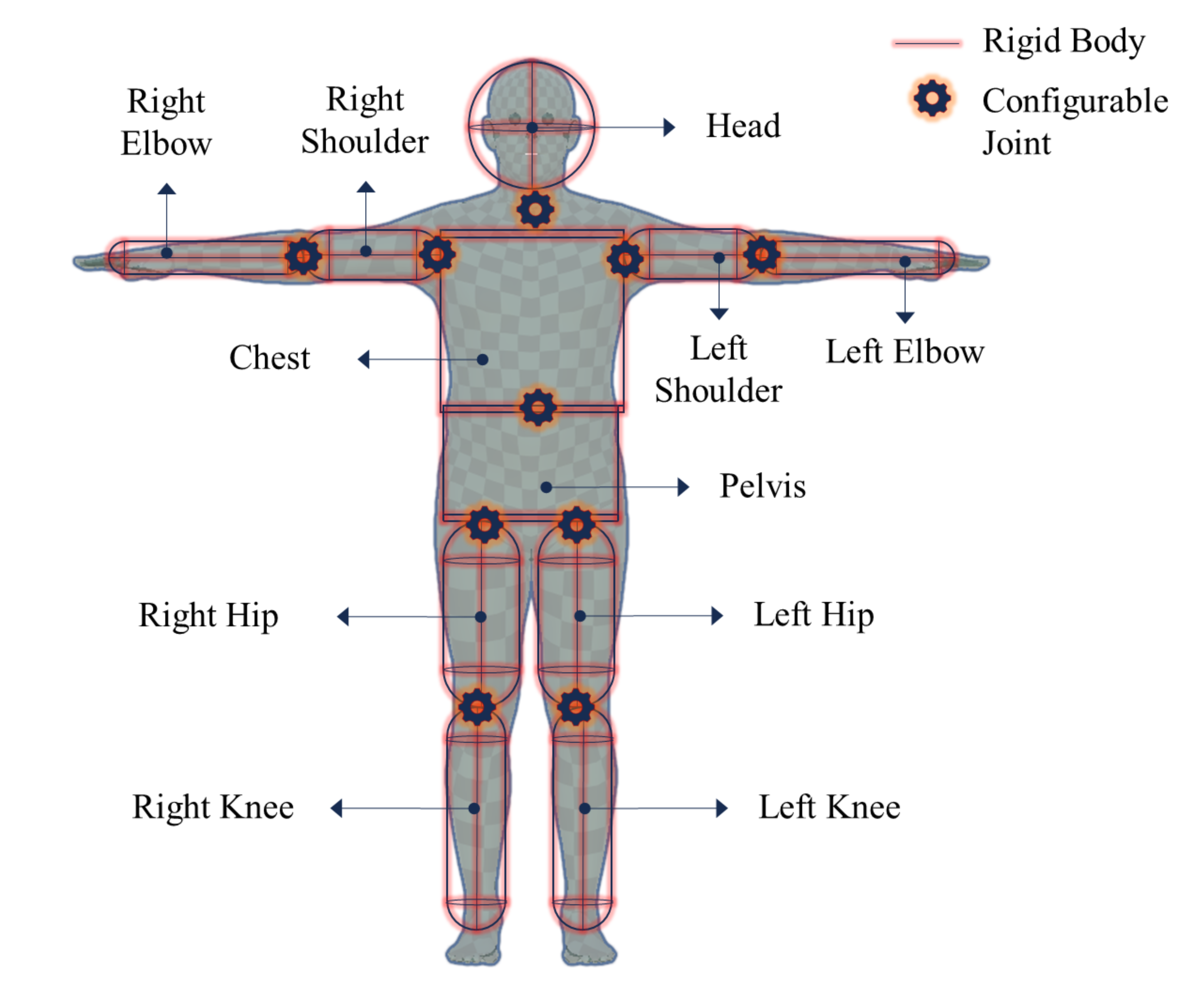}
  \caption{Ragdoll representation of an avatar with rigid bodies.
This diagram shows the structure of a ragdoll created using primitive-shaped rigid bodies and configurable joints.
  }
  \vspace{-0.5cm}
  \label{fig:ragdoll}
\end{figure}

\subsection{Physics-based Model}
%We employed a basic feedback control system to simulate our physics-based avatars. 
% These avatars not only follow physical constraints, but the presence of their responsive nature is crucial for creating immersive virtual experiences \cite{llobera2013telling, llobera2021interactive}.
We followed an approach similar to other recent work \cite{vyas2023exploring, alvarado2022generating}, which utilizes two types of characters: a dynamic character, often referred to as a ragdoll, and a kinematic character. The dynamic character was constrained to follow the motion of the kinematic character, which acted as a target. As illustrated in Figure \ref{fig:ragdoll}, the ragdoll was constructed from cuboid, cylinder, and sphere-shaped rigid bodies. This humanoid skeleton, made up of rigid bodies, was constrained to follow the target reference motion through torques and forces applied to the joints.

% The Proportional-Derivative (PD) controller, a feedback control system, is commonly used to control the joint activation of dynamic characters when kinematic reference motions are available \cite{zordan2002motion, yin2007simbicon, silva2021topple}. 
The feedback control system used a Proportional-Derivative (PD) Controller, which is commonly used to control the joint activation of dynamic characters \cite{yin2007simbicon, silva2021topple}. We used the Unity\texttrademark{} game engine for character simulation. The ragdoll consisted of a skeletal structure with 22 degrees of freedom (DoF) and 10 joints. The DoF in each joint represents the relative translation or rotation of body segments through prismatic or revolute joints using the built-in configurable joints in Unity.
The PD controller acted as a servo for each degree of freedom at each time step:
\[
\mathbf{\tau_{t}} = -k_{p}(\mathbf{q} - \mathbf{\Bar{q}}) - k_{d}{\dot{q}}
\]

where $\mathbf{\Bar{q}}$ represents the desired pose, $\mathbf{q}$ denotes the current pose, and $\dot{q}$ is the joint velocity. We used the mass distribution for rigid bodies and used gain values for controllers similar to those mentioned by Vyas et al. \cite{vyas2023exploring} to achieve optimal control.

\subsection{Stimuli}
%For this experiment, we used physics-based virtual avatars as stimuli.
Stimuli consisted of animated scenes, each populated with a small-scale crowd consisting of twelve physics-based avatars. We selected twelve male and twelve female subjects from the dataset captured by Hoyet et al. \cite{hoyet2013evaluating}. This dataset includes marker data from 40 subjects, along with their anthropometric measurements. 
% The dataset features various walking styles, including those performed with a metronome (indicating a steady pace) and without a metronome (allowing subjects to choose their own walking pace). We opted for motions without a metronome to introduce natural variability, reflecting the diversity typically seen in a group of people. 
Since body shape was the primary focus of our study, we selected subjects based on their variability in body weight and height. Specifically, within each BMI group (low, average, and high), we chose four subjects, ensuring variability in their body height as well. This selection was consistent for both male and female stimuli, resulting in a diverse set of body shapes and motions.

To create the body models, we used methods similar to those described in previous works \cite{russell2023detection, vyas2023exploring}. The MoSh++ method \cite{mahmood2019amass} was applied to the marker motion data to convert it into the SMPL mesh \cite{pavlakos2019expressive, romero2022embodied}. This method generated a 3D surface with 10,777 triangles, which was then converted into a $C\#$ readable format using the bmlSUP player \cite{bebko2021bmlsup} for animation playback in Unity.

\begin{figure}[t]
  \centering
  \includegraphics[width=1.0\linewidth]{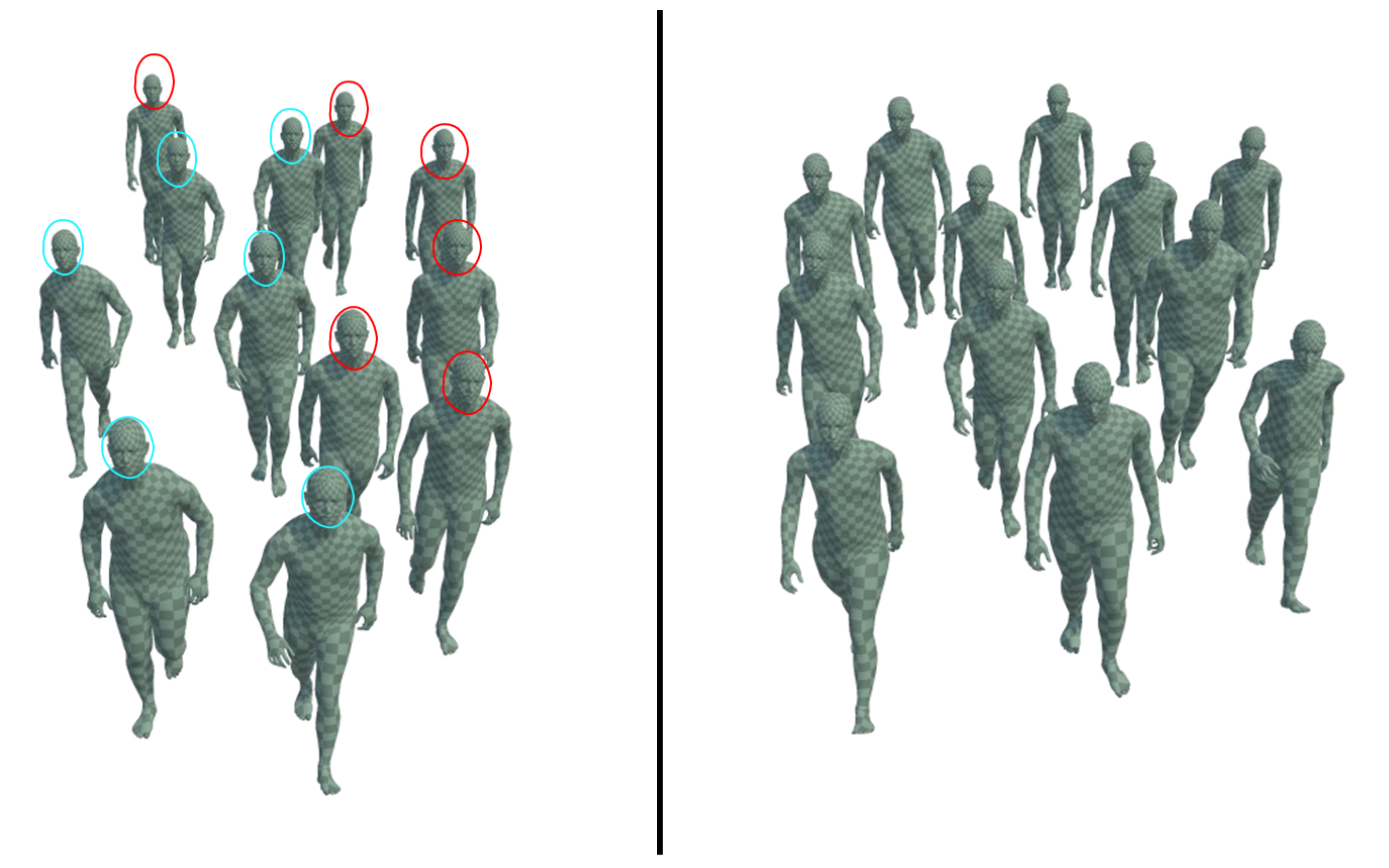}
  \caption{Stimulus Creation:
\textit{Motion Clone} (left): Featuring characters with two distinct motions,  the first circled in green, and the second in red (out-of-step motions). Each character has one of six different body shapes. \textit{Baseline} (right): All characters have unique body shapes and unique motions.
  }
  \vspace{-0.1cm}
  \label{fig:stimuliCreationCrowd}
\end{figure}

The animation clips were rendered in Unity within an empty white background environment. We prioritised the realism of motion and shape, rather than on detailed skin and clothes rendering, which could be a confounding factor. 
% To avoid the uncanny valley effect \cite{mori2012uncanny, schwind2018avoiding},
A checkerboard green texture was applied to the avatars, as seen in Figure \ref{fig:stimuliCreationCrowd}. The camera angle was kept consistent across all rendered clips, ensuring that all twelve virtual avatars were visible at any given time. We chose the aerial viewpoint for the camera similar to other studies on crowd perception \cite{daniel2021perceptually, adili2021perception, kulpa2011imperceptible, pravzak2011perceiving}.
%We introduced position offsets in each scene because it is possible that participants might have focused on just a few virtual avatars, making changes in body shape easily noticeable. 
We also ensured that no two clips featured virtual avatars in the same location, thus mitigating any positional bias.\\

%\subsubsection*{Baseline}
{\raggedleft{\textbf{Baseline:}}}
The baseline scenes consisted of virtual avatars where each one had a unique motion and body shape.
% The interplay between motion and body shape was critical, as the perception of a character's motion is influenced by its appearance, and certain motions may look different depending on the body shape performing them.
% As previous research has shown, humans are sensitive to even small discrepancies in virtual simulations \cite{russell2023detection}.
To avoid bias, we randomized the combinations of body shapes and motions in the stimuli. None of the virtual avatars in the scene retained their original motion and body shape pairing, thereby ensuring a fair comparison for motion clones.
The Fisher-Yates shuffle algorithm was used to randomize the lists of motions and body shapes. 
For each sequence, the shuffled list of motions was paired with the shuffled list of body shapes to create the twelve crowd characters. Unique sequences were generated by permuting the shuffled lists, and we ensured that repetitions were avoided by tracking previously used sequences.\\
% {\color{red}The experiment was designed to distribute all actors equally among the possible motions. The body shapes were divided into BMI sub-groups, reducing potential bias.}

%\subsubsection*{Motion Clone}
{\raggedleft{\textbf{Motion Clone:}}}
The motion clone scenes featured 12 virtual avatars with one or more motions applied to them. We explored four levels of motion cloning: 1, 2, 3 ,and 6 motions evenly distributed across the avatars. Additionally, we included different levels of body shape variation: 1, 3, 6, and 12 body shapes, similarly distributed. Therefore, each scene varied the number of unique motions and the diversity of body shapes among the twelve virtual avatars.

To avoid bias from specific motion and body shape pairings, we randomized the avatars' body shape, motion, and position in each scene. This approach ensured that each stimulus presented a unique combination of motions and body shapes.
We ensured that repeated motions (due to cloning) were out-of-step with each other to prevent these motion clones being too easy to detect. This resulted in a unique set of combinations for each stimulus.

\begin{figure}[t]
  \centering
  \includegraphics[width=1.0\linewidth]{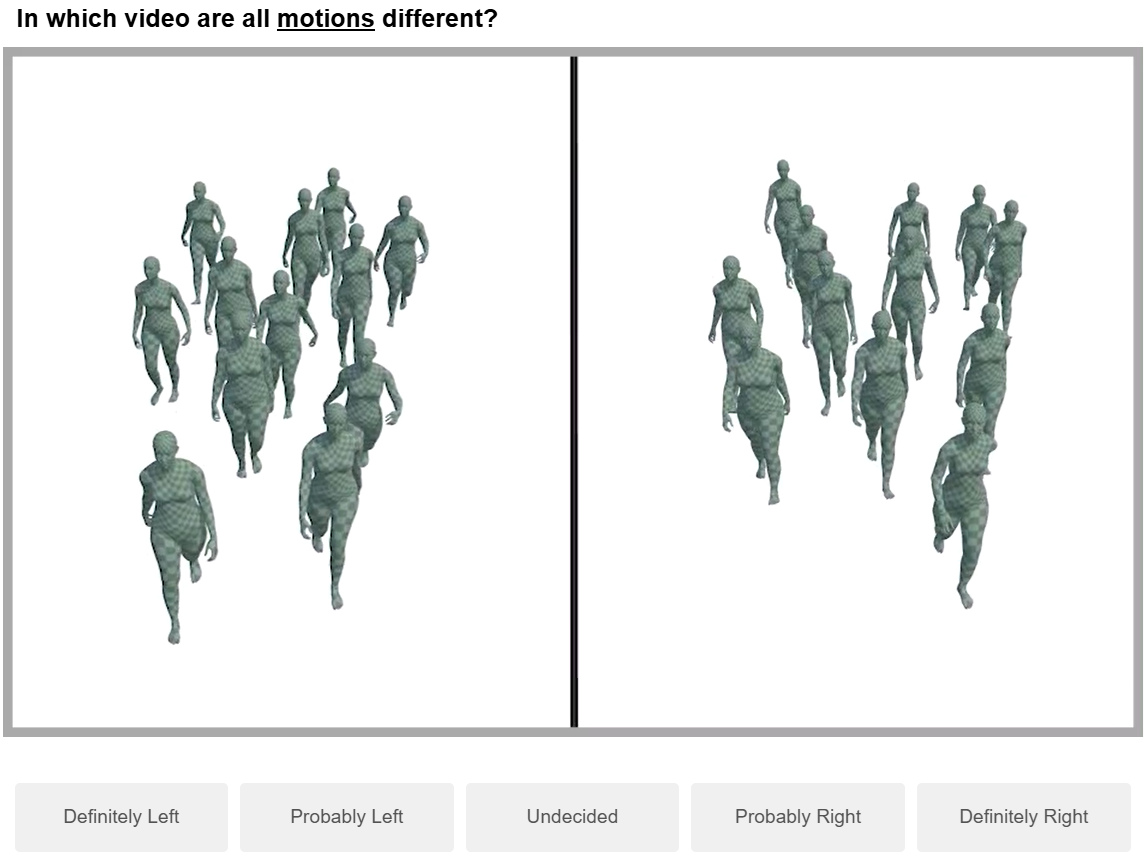}
  \vspace{-0.3cm}
  \caption{Example Stimulus: Participants indicated their confidence level when selecting the side on which all motions were unique.
  }
  \label{fig:stimuliCrowd}
  \vspace{-0.25cm}
\end{figure}

\subsection{Method}
Twenty-two participants (9F, aged 18-60+) from diverse backgrounds were recruited
%to perform the experiment online. Participants were recruited
through internal mailing lists and social media platforms. 
%None of these participants had been involved in the motion capture sessions used for stimuli creation. 
The experiment was run entirely online and participation was entirely voluntary and anonymous. After providing informed consent and demographic information, 
%and withdraw participation at any time during the experiment.
%\subsection{Procedure}
participants were presented with pairs of video clips showing crowd scenes side by side on their computer screens. The baseline scene with all unique avatar motions was presented on one side, while on the other side, a scene with cloned motions was displayed. Participants were tasked with choosing the side where they believed all motions were different. Essentially, this was a clone detection task, as failure to detect them would result in a motion clone scene being erroneously chosen.

Participants were assigned to one of two groups based on the sex of the avatars in a counterbalanced manner. The experiment therefore employed a mixed design to investigate the effects of the between-subjects factor \textit{SEX} (Male or Female avatars) and the within-subjects independent variables BODY (body shape level) and MOTION (motion clone level) on the detection of motion clones.  Within each \textit{SEX} group, there were combinations of 4 BODY levels (1, 3, 6, 12 bodies) $\times$ 4 MOTION levels (1, 2, 3, 6 motions), resulting in a total of 16 clip combinations.

%As in the stimuli presentation, one side of the screen displayed the Baseline clip, while the other side displayed one of the motion clone clips. 
To control for side bias (left or right), the Baseline clip was shown on the left side for half of the motion clone clips (16 clips) and on the right side for the other half (16 clips). Participants were asked to click on the screen to start the videos, and both videos played simultaneously, allowing them to control when the video would start, but they could not pause or replay it. 

This design ensured that participants’ choices were based on their initial impressions and prevented them from intentionally finding patterns by pausing or replaying the videos. As already mentioned, to eliminate any familiarity effects related to the position or combination of motion and body shapes, the 32 motion clone clips had unique combinations for both sides. Within each \textit{SEX} group, the video pairs were randomized for each participant.
At the start of the experiment, participants were shown example stimuli featuring crowds with either different motions or identical motions. This was followed by a series of practice questions to familiarize participants with the task. These examples featured actors with body shapes and motion combinations similar to the actual stimuli, but none of these actors' body shapes and motions were used in the main experiment.

The experiment presentation consisted of 16 clips with the Baseline on the left and 16 with the Baseline on the right, with each set being repeated twice. This resulted in a total of 64 questions. Each video clip lasted 6 seconds. Participants were asked to watch the video clips and select the side on which they believed all motions were different. They answered using a 5-point Likert scale, allowing them to express their confidence level: 'Definitely left', 'Probably left', 'Undecided', 'Probably right', and 'Definitely right'. An example stimulus for the experiment is shown in Figure \ref{fig:stimuliCrowd}. After selecting an answer, participants pressed the 'Next' button, which only appeared after the clip had finished playing. The total duration of the experiment was approximately 12-15 minutes on average.

\begin{figure}[t]
  \centering
  \includegraphics[width=1.0\linewidth]{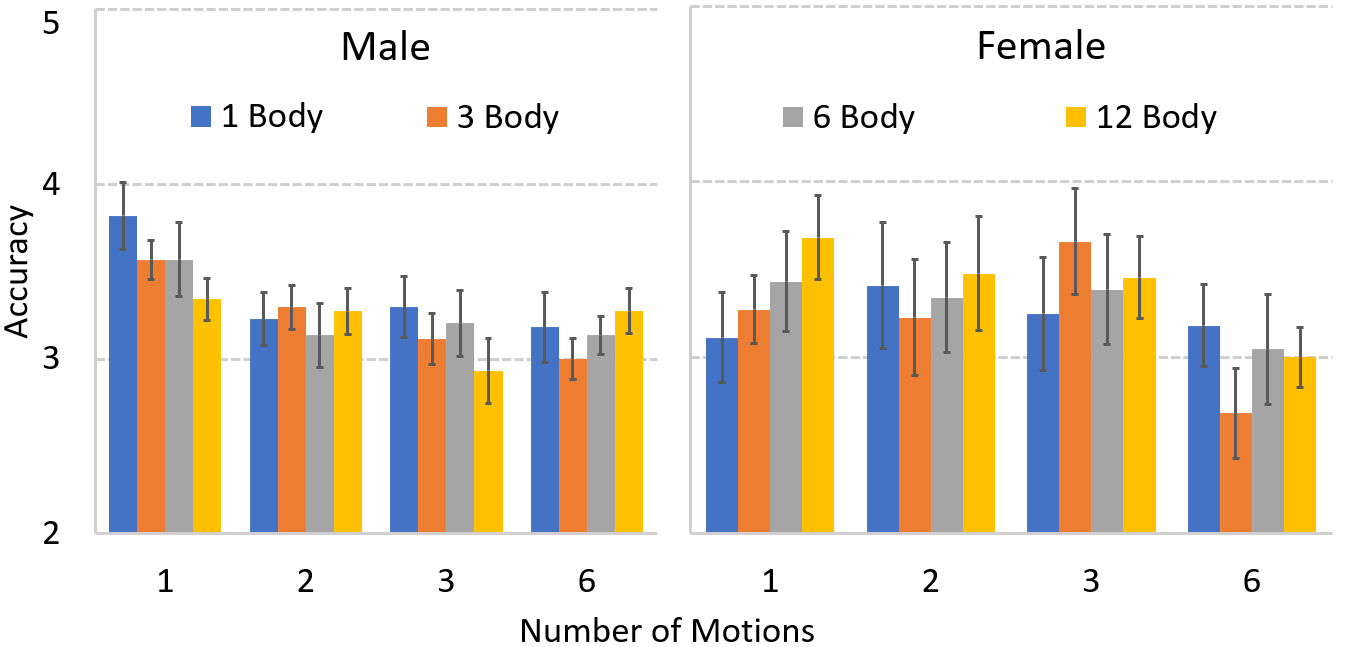}
  \vspace{-0.3cm}
  \caption{Mean participant responses (with standard error bars) showing accuracy for identifying the gold standard crowd. Higher values indicate more accurate motion clone detection.}
  \vspace{-0.5cm}
  \label{fig:Meanresults}
\end{figure}

\section{Results} 
We first converted all participants' 
5-point Likert scale responses to accuracy rates. If the participant correctly chose the side with the gold standard, they received 5 (for definitely) or 4 (for probably). They received 3 for undecided, and 2 or 1 if they chose incorrectly (for probably and definitely respectively).
Figure \ref{fig:Meanresults} shows the mean accuracy and standard errors for participants' ability to identify the gold standard crowd across varying combinations of motions (1, 2, 3, and 6) and body shapes (1, 3, 6, and 12) for Male and Female stimuli. Participants were asked to choose the side where they believed all the motions were different. The higher values therefore also indicate a higher motion clone detection rate, whereas lower accuracy rates mean that motion clones were masked more effectively by the bodies.

No consistent relationship can be observed between body shape diversity and motion clone detection across both groups. The highest accuracy/clone detection rate (3.8) overall was achieved for one motion and one body shape with the male stimuli. For males, all detection rates for one motion were at above chance level (accuracy of 3, indicating undecided), whereas for 2 or more motions, participants appear to have been guessing. Furthermore, the error bars for one motion with one body vs. with 12 bodies do not overlap, suggesting that, at least in the case of the male stimuli, the body shape diversity helped to mask the motion clones.
\newpage

  The lowest accuracy (2.7) overall was observed with six motions and three body shapes for the female stimuli, which is consistent with our hypothesis that increasing motion variety will reduce clone detection performance.
  However, for the females, the highest accuracy (3.7) was achieved when there was one motion and 12 body shapes, which is not consistent with the body shape results for the male stimuli. It should be noted that the standard errors for the females are higher than for the males, indicating that there was less agreement between the participants in that group. The smaller standard errors for the male stimuli indicate that participant performance within that group was more consistent. 

\begin{table}[t]
\begin{center}
%\vspace{-0.3cm}
\caption{\small ANOVA showing main and interaction effect tests with effect sizes ($\eta^{2}$) and significance values ($p$).}

\label{table:statsCrowd}
\small
\begin{tabular}{lllllll}\hline\hline
&\textbf{Effect Tested}  & & \textbf{dof}
&\textbf{F-Test} & \textbf{$\eta^{2}$} & \textbf{p}\\\hline\hline
%\multicolumn{3}{ l }{\rule{0pt}{3ex}\emph{Tests of Main Effects}}\\  	\hline

&  MOTION  & & 2.5, 49.2	 & 3.05 & 0.132 & 0.046\\
&  BODY  & & 3, 60 & 0.47 & 0.023 & 0.701\\
&  SEX  & & 1, 20	 & 0.01 & <0.001 & 0.937\\

%\multicolumn{3}{ l }{\rule{0pt}{3ex}\emph{Tests of Interaction Effects}}\\  	\hline

&  MOTION x BODY  & & 9, 180	 & 0.73 & 0.035 & 0.682\\
&  MOTION x SEX  & & 2.5, 49.2	 & 1.56 & 0.073 & 0.216\\
&  BODY x SEX  & & 3, 60	 & 1.72 & 0.079 & 0.172\\

%\multicolumn{3}{ l }{\rule{0pt}{3ex}\emph{Teswo-way Interaction Effects}}\\  	\hline

&  MOTION x BODY x SEX  & & 9, 180 & 1.41 & 0.066 & 0.188\\\hline
\end{tabular}

\end{center}
\vspace{-0.7cm}
\label{tab:Stats}
\end{table}

A repeated measures mixed ANOVA was conducted on the mean accuracy rates, with between-subjects factor SEX (M/F) and within subjects variables MOTION (1/2/3/6) and BODY (1/3/6/12). The results are shown in Table \ref{tab:Stats}.
%For detailed results, see Appendix A (Table \ref{table:statsCrowd}).
The MOTION variable violated sphericity, so the degrees of freedom were corrected (Hyunh-Feldt). 
The only significant effect found was for MOTION ($F(2.5, 49.2)=3.05, p<0.05)$, with an effect size of 0.132. This provides some evidence that, as the variety of motions increased, participants were more likely to detect differences between the virtual avatars. Post-hoc pairwise comparisons (Bonferonni) showed that this effect can be explained by the highest accuracy with one motion, and the lowest with 6 motions. Albeit a small effect, it indicates that greater motion diversity helps to reduce the detection of motion clones, thereby making the virtual crowd scene appear more natural.

Although we did not find any significant effect of body shape diversity, indicating that avatar body shapes did not strongly affect participants' ability to distinguish between unique and cloned motions overall. This may be partially explained by the more inconsistent results for female avatars. The Male results in Figure \ref{fig:Meanresults} show that the highest accuracy recorded was for 1 body and 1 motion, where accuracy was above chance level. 

%This finding indicates that visual elements like body shape alone may not be sufficient to change how people perceive motion realism in crowds. It also implies that while body shape diversity is important for making avatars visually distinct, it may not play a major role in detecting cloned movements compared to motion variety.

%Similarly, the sex of the virtual avatars did not significantly impact clone detection. This means that whether the avatars were male or female had no influence on how participants identified motion clones. There were also no significant interaction effects.

%Overall, the results suggest that motion variety is far more important than body shape or sex in detecting motion clones in virtual crowds. While body shape and sex help with the overall visual realism, they do not significantly affect how people perceive motion clones. 
%Motion diversity, on the other hand, plays a critical role in creating believable virtual environments.

\section{Discussion}  
In this study, we explored the impact of body shape diversity on the detection of motion clones in the context of small-scale virtual crowds. We also created dynamically controlled virtual avatars for the stimuli, which allowed motion to be consistently applied across all body shapes. 
We first hypothesized that increasing motion variety would reduce the detection rate of motion clones, and we did find evidence to show that this was indeed the case. Previous research has also highlighted the importance of motion variation in crowd simulations \cite{adili2021perception}. Therefore, motivated by previous work \cite{russell2023detection, vyas2023exploring} that explored the relationship between body shape and motion for virtual avatars, we hypothesised that increasing body shape diversity would make the detection of motion clones more difficult. The experiment was designed to simulate a variety of body shapes and motions, with virtual avatars categorized by Body Mass Index (BMI) and varying levels of motion and body diversity. 

The significant, albeit small, main effect of motion variety showed that increased diversity in motions helps to reduce the detection of motion clones. In fact, above chance detection levels were only perceived for one motion in the case of the male stimuli, with other motion levels at chance. 

\newpage
\noindent
The performance of the participants in the female group was more variable, which may have reduced the size of the effect, but the lowest accuracy rate was found for 6 motions with the female stimuli, further reinforcing the case for accepting the motion variety hypothesis. These results are consistent with those of Prazak et al. \cite{pravzak2011perceiving}, who found that participants could not tell the difference between a crowd with all unique motions and one with three evenly cloned motions.

Contrary to our second hypothesis, body shape diversity did not significantly affect participants' ability to detect motion clones. This aligns with previous findings by Hoyet et al. \cite{hoyet2016perceptual}, who found that motion variety had a greater impact on crowd perception than other visual elements. Thus, although variations in body shape contribute to the overall visual diversity of avatars, increasing the number of motions to two or more may be sufficient to give the illusion of a fully varied crowd.

\begin{figure}[h]
  \centering
  \includegraphics[width=1\linewidth]{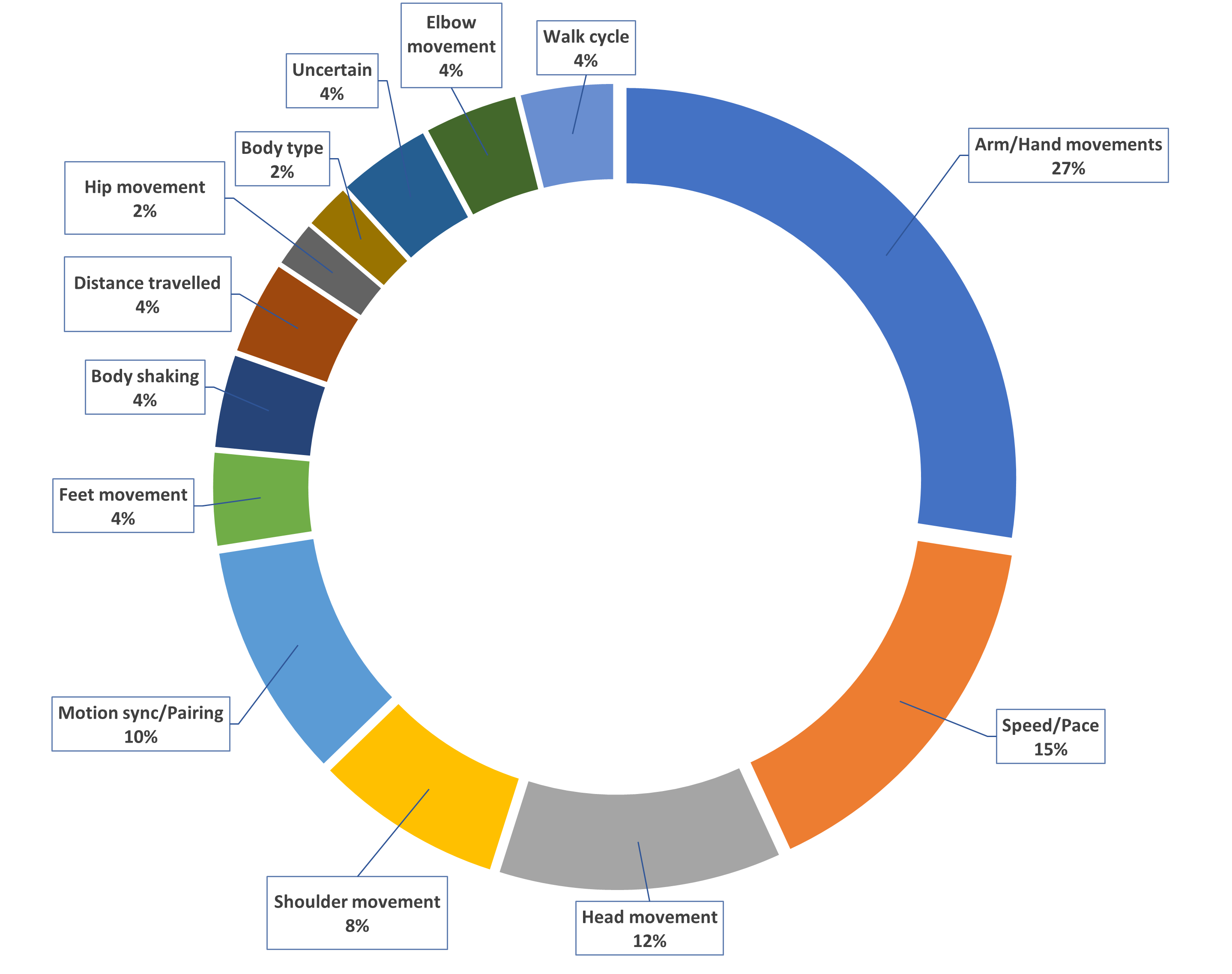}
  \caption{Distribution of key factors considered by participants during experiment.
  }
  \vspace{-0.3cm}
  \label{fig:CrowdComments}
\end{figure}

We asked participants to provide some comments on what basis they made the decisions at the end of the study. Figure \ref{fig:CrowdComments} illustrates the distribution of some key factors mentioned in the comments. "Arm/Hand movement" emerged as the primary factor. This could be attributed to a particular motion in the female stimuli with a distinct arm swing. This observation could also explain why performance was worse for six motions than for one motion in the female stimuli group. It would be worth investigating the role of distinctiveness in future such studies.
Other important factors included "Speed," "Head movement," and "Shoulder movement." These comments may help to guide the design of better experiments and crowd scenes.

Future work could explore additional variables, such as clothing or facial features, which may interact with motion to influence perceived crowd realism. Exploring larger crowds and different interaction scenarios may also reveal other effects of body shape. Perception of secondary motion, such as subtle environmental reactions, would also be an interesting direction to explore.  
% Additional variables, such as clothing or facial features, may also interact with motion to affect perceived crowd realism, while testing larger crowds and different interaction scenarios may reveal other effects of body shape. Perception of secondary motion, such as subtle environmental reactions, would also be an interesting direction to explore. 
In conclusion, this study contributes to a deeper understanding of how body shape and motion clones impact crowd perception, suggesting that motion variety plays a more important role than body shape in influencing realism in small-scale virtual crowds. Further studies are needed to test the limits of motion clone detection and thereby inform more efficient crowd simulations that balance realism with computational demands.


\acknowledgments{
This work has received funding from the European Union’s Horizon 2020 research and innovation programme under the Marie Skłodowska Curie grant agreement No 860768 (CLIPE project).
\vspace{-0.25cm}}

\bibliographystyle{abbrv-doi}

\bibliography{Paper}
\end{document}